\documentstyle[12pt,epsf,epsfig,wrapfig]{article}
\begin{document}
\begin{center}
{\large \bf
    QCD Aspects of Polarized Hard Scattering Processes
    \footnote{ To appear in the Proceedings of the 12th International
        Symposium on High-Energy Spin Physics Amsterdam from September
        10--14, 1996 (World Scientific).
    }
\\ }
\vspace{5mm}
    John C. Collins
    \footnote{E-mail: {\tt collins@phys.psu.edu}.}
\\
\vspace{5mm}
{\small\it
    CERN---TH division, CH--1211 Geneva 23, Switzerland\\
        (Permanent address:
        Penn State University,  104 Davey Lab., University Park
        PA 16802, U.S.A.)
\\ }
\end{center}

\begin{center}
                            ABSTRACT

\vspace{5mm}
\begin{minipage}{130 mm}
\small
    I give an overview of what our present knowledge of QCD
    predicts and does not predict for polarized hard scattering.
    For experimental programs, a big issue is how much further we
    can expect our theoretical understanding of QCD to improve.
\end{minipage}
\end{center}

\section{Introduction}

In the area of of high-energy spin physics, there are many
experiments that are in various stages of construction and
proposal whose data will need to be analyzed.  They will cover a
much wider range of phenomena than previous experiments.
Therefore, I will review what we know from QCD about hard
processes, what we don't know (at least not yet), and the areas
in which it is realistic to expect our knowledge to improve.

\section{State of QCD}

Our ability to make predictions from QCD is highly conditioned by
its asymptotic freedom.  Thus perturbation theory can be used to
make useful predictions for processes governed by short-distance
phenomena. For non-perturbative infra-red phenomena, the only
currently available methods for making predictions from first
principles are lattice Monte-Carlo calculations. However, these
only provide results in Euclidean (imaginary) time, and so are
only useful for static quantities like masses.

Scattering processes combine short- and long-distance phenomena,
and calculations are based on use theorems about the asymptotics
of amplitudes and cross sections. Thus we have ``factorization
theorems'' for processes with a hard scattering (deep-inelastic
scattering, jet production, etc), where a cross section is a
product of a non-perturbative and a perturbative factor.  The
well-known Monte-Carlo event generators result from a
particularly complicated (but approximate) case of these
theorems.

\subsection{Factorization}

Now I will review the features of
a typical factorization theorem [1].
Illustrated in Fig.\ 1, is the
one for a deep-inelastic structure function.  The lines in the
upper part of the graph are far off-shell, while those in the
lower part of the graph form a single-particle density (called a
parton distribution function).  The final-state lines in the
upper part of the graph can be treated as effectively off shell
in the context of a sufficiently inclusive cross section that the
details of the final-state are not resolved.

\begin{wrapfigure}{R}{6cm}
    \epsfig{figure=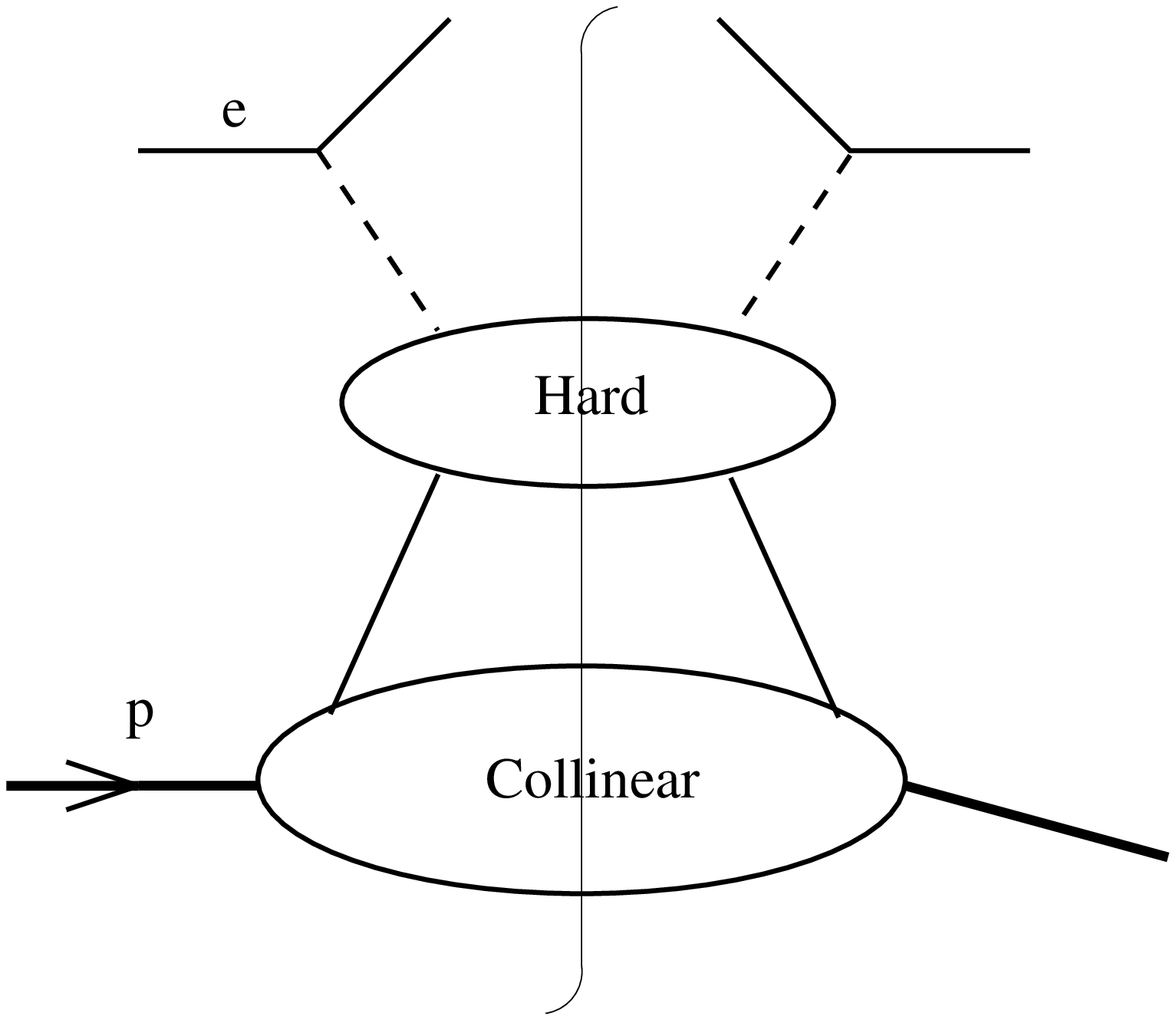,width=6cm}\\
    {\small Figure 1: Factorization for deep-inelastic structure
        function.}
\end{wrapfigure}

The corresponding formula is
\begin{eqnarray}
   F_{1}(x,Q) &=& {\rm pdf}(x) \otimes \mbox{``Wilson coefficient''}
\nonumber\\
    && + \mbox{power-suppressed terms}.
\end{eqnarray}
This is a provable impulse approximation, with the parton
distribution being a function only of a longitudinal momentum
fraction, because of the relativistic kinematics.
The hard scattering coefficients (``Wilson coefficients'') are
perturbatively calculable in powers of $\alpha _{s}(Q)$.
The parton densities can (in principle) be measured in a few
experiments and then used to predict other processes that have a
factorization theorem.

The non-trivial features of factorization are the need for
higher-order corrections to the coefficient functions and the
DGLAP evolution of the parton densities.

\subsection{Spin}

\begin{wrapfigure}{R}{5cm}
    \epsfig{figure=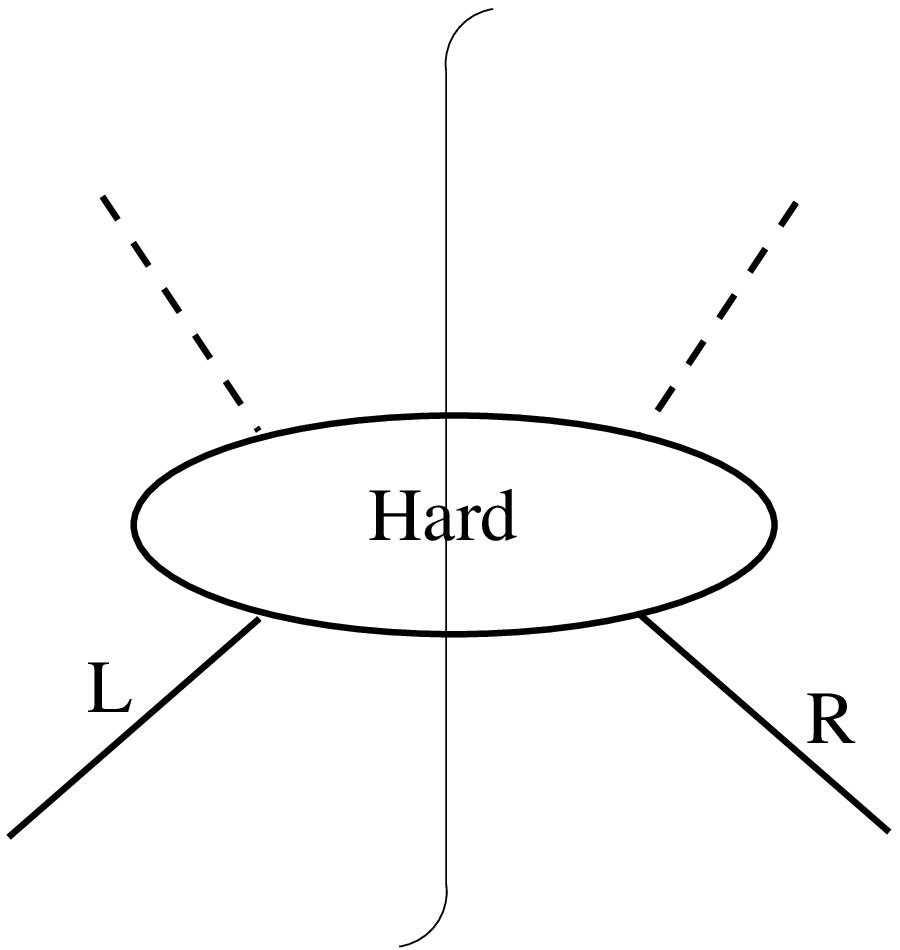,width=5cm}\\
    {\small Figure 2: Interference between amplitudes with left
       and right handed quarks is zero for perturbative
       hard-scattering coefficients.
    }
\end{wrapfigure}

When we treat polarized processes, simple generalizations of the
same factorization theorems continue to be provable [2].  The
complication is that the parton lines entering the hard
scattering need to be equipped with helicity density matrices.
The combination of parton densities and the density matrices can
be conveniently represented in terms of the unpolarized parton
densities and some spin-asymmetry densities.  For a spin-half
hadron (like a proton), we have longitudinal spin asymmetries
($\Delta u$, $\Delta d$, \dots, $\Delta g$) and transverse spin asymmetries (or
`transversity' densities, $\delta u$,
etc).  Jaffe has used the notation $h_{1}$ for the transversity
densities, but I prefer the notation $\delta $ or $\delta _{T}$.  Because the
gluon has spin 1, it can be proved from rotation invariance that
there is no transversity asymmetry for the gluon in a spin-half
hadron.

A particularly important set of predictions arises because QCD
predicts that chirality is conserved in hard scattering
coefficients.  This implies that there is no interference between
amplitudes for left and right-handed quarks (Fig.\ 2).
Thus many transverse spin asymmetries are zero in the leading
twist-2 approximation.  A typical case [3] is $g_{2}$ in DIS.  The
phenomenology of higher twist processes is much more difficult.

In full QCD, including its non-perturbative part, chirality
conservation is broken both by quark masses and by the
spontaneous symmetry breaking that gives the pion its small mass.
This breaking is relevant for parton densities (and fragmentation
functions) but not for the coefficient functions.

\subsection{Status}

The factorization theorems are established [1] for many processes
to all orders of perturbation theory (and not just to the leading
logarithm approximation).  A certain amount of intuition together
with some non-perturbative parts of the proofs indicate that the
theorems are valid more generally.  The primary difficulties in
establishing the theorems and generalizing them are the intricate
cancellations of initial- and final-state interactions that is
necessary to avoid correlations between the hadrons in
hadron-hadron collisions.

\begin{sloppypar}
Typical processes for which we have factorization theorems are:
\begin{itemize}

\item
    DIS (deep-inelastic scattering): inclusive.

\item
    DIS: semi-inclusive, production
    of jets, heavy quarks, etc.

\item
    Drell-Yan, i.e., hadron-hadron to high mass muon pairs, etc.

\item
    Hadron-hadron to jets and high $p_{T}$ hadrons.

\item
    Hadron-hadron to direct photons at high $p_{T}$.

\item
    Hadron-hadron to heavy quark inclusive.

\item
    $e^{+}e^{-}$ to jets, etc.

\end{itemize}

In addition, there are theorems on elastic scattering [4], but
with complications that I will review in the next section.

\end{sloppypar}

\section{What don't we know? (At present, from theory)}

First, we do not know how to obtain the parton densities from
first principles, except for certain moments that correspond to
conserved Noether charges. Hence the parton densities must be
obtained from experiment, with the aid of the perturbatively
predicted hard scattering coefficients. However, the evolution of
the parton densities is predicted perturbatively.

Identical remarks apply to fragmentation functions.

Another area of uncertainty is higher twist physics, that is,
power-law corrections to a normal (``twist-2'') factorization
theorem that obeys dimensional counting rules for its
$Q$-dependence.  (By $Q$, I mean a measure of the scale of the
hard scattering.)  To some extent there are real theorems of the
factorization type, at least for twist-3 and twist-4.

The difficulties arise in two areas.  First, it
is hard to separate a non-leading power from the leading power,
in view of the logarithmic corrections to the leading power.
This problem does not apply to observables that are zero at the
twist-2 level; such observables are common in transverse
polarization asymmetries.  Cases are $G_{2}$ in DIS and the
single-spin asymmetry in high $p_{T}$ particle production
($p+p\to \pi +X$).

Even when one can extract the higher twist observable, it is hard
to analyze phenomenologically.  The difficulty is that the
cross-section is expressed in terms of non-perturbative
quantities that in the case of higher twist involve things like
quark-gluon correlations in a hadron.  Integrals over
longitudinal momentum fractions are involved, which make it
difficult to extract, for example, a correlation function
$C(x_{1},x_{2})$ as a function of $x_{1}$ and $x_{2}$.  At the present state
of the art, we must treat such non-perturbative functions as
unknown theoretically and only obtainable by analysis of
experiments.

Of course, there has been much work in these areas, but the
important point is that it is hard to do really crisp
phenomenology.

There are also results [4,5] for elastic scattering at large $t$.
But these results, from the phenomenological point of view,
suffer from the same disadvantage as higher-twist quantities, of
involving integrals over light-cone wave functions.  Again, it is
very hard to extract the wave functions unambiguously from
experiment for that precise reason.  In addition, the correct
form of the factorization theorem is not so simple, with a
combination of different mechanisms: In addition to the pure
short distance process [4], there is the Landshoff process [5],
with its Sudakov suppression.

I do not want to minimize the amount of good work that has been
done.  But in view of the large effort needed to make twist-2
phenomenology precise, I tend to blanch at what is needed to do
corresponding work for higher twist and for (high $t$) elastic
scattering.

\section{Fragmentation and quark polarimetry}

Another area of unknown quantities is that of fragmentation.  The
fragmentation functions, i.e., the distribution of hadrons
in the fragmentation of partons, are less widely discussed than
that parton densities, but are of approximately equal status
theoretically.  There has been useful phenomenology of the
unpolarized case [6], but the polarized case is almost {\it terra
incognita}.

The basic idea is given by considering semi-inclusive DIS, for
which the parton model is summarized in Fig.\ 3. The full QCD
factorization differs only by having higher order corrections to
the hard scattering and needing evolution of the parton densities
and fragmentation functions.  The process is one like
$e+p \to  e+\Lambda +X$ or $e+p \to  e+\pi \pi +X$,
where one considers the production of a
system of one or more hadrons away from the beam fragmentation
region.  We have a theorem of the form
\begin{equation}
   \sigma  = \mbox{pdf}
       \otimes \mbox{hard scatter}
       \otimes \mbox{fragmentation} ,
\end{equation}
for the leading power.

\begin{wrapfigure}{R}{6cm}
    \epsfig{figure=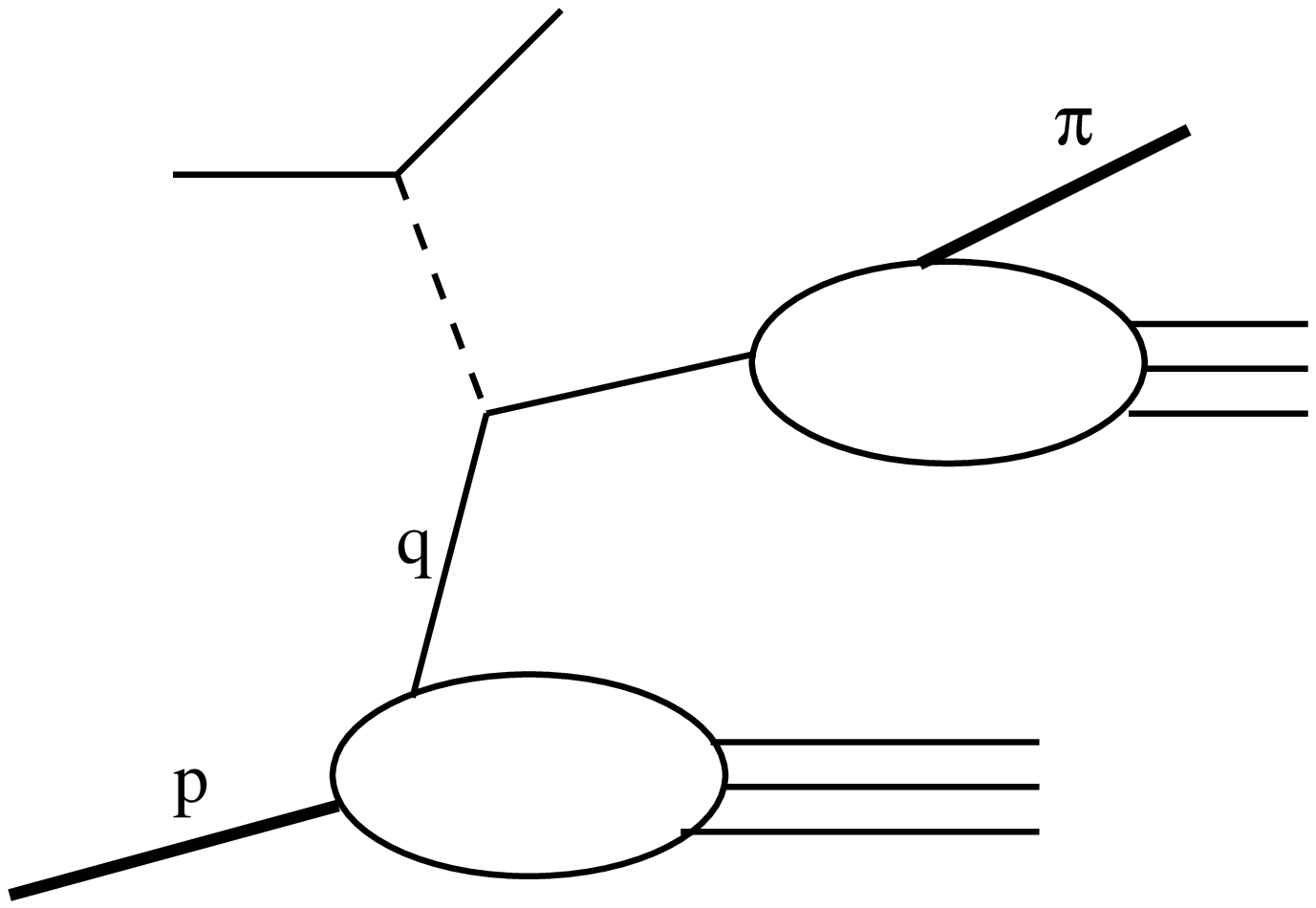,width=6cm}\\
    {\small Figure 3: Parton model for semi-inclusive DIS.}
\end{wrapfigure}

A number of people [7--9], including myself, have worked on this
subject recently.  We consider the concept of measuring the
polarization state of a quark or a gluon to be both fundamental
and interesting.

There are at least three measurements that have been proposed:
\begin{itemize}

\item Measure the polarization of a $\Lambda $: $q \to  \Lambda  + X$.
    Data has recently become [10] available from the ALEPH
    collaboration for the case of longitudinal polarization.
    They indicate a large polarization transfer (tens of
    percent) at large $z$ --- see Fig.\ 4.

\item Handedness of jets [8], for measuring the helicity of a
    quark or gluon.

\item The azimuthal distribution of hadrons around a jet axis,
    for measuring quark transverse polarization.[9]

\end{itemize}
The last two were reviewed by Efremov in his talk here [11],
particularly as regards the experimental situation, where data in
$e^{+}e^{-}$ begins to show a possible non-zero effect, at present of
marginal significance.

\begin{wrapfigure}{R}{8cm}
    \epsfig{figure=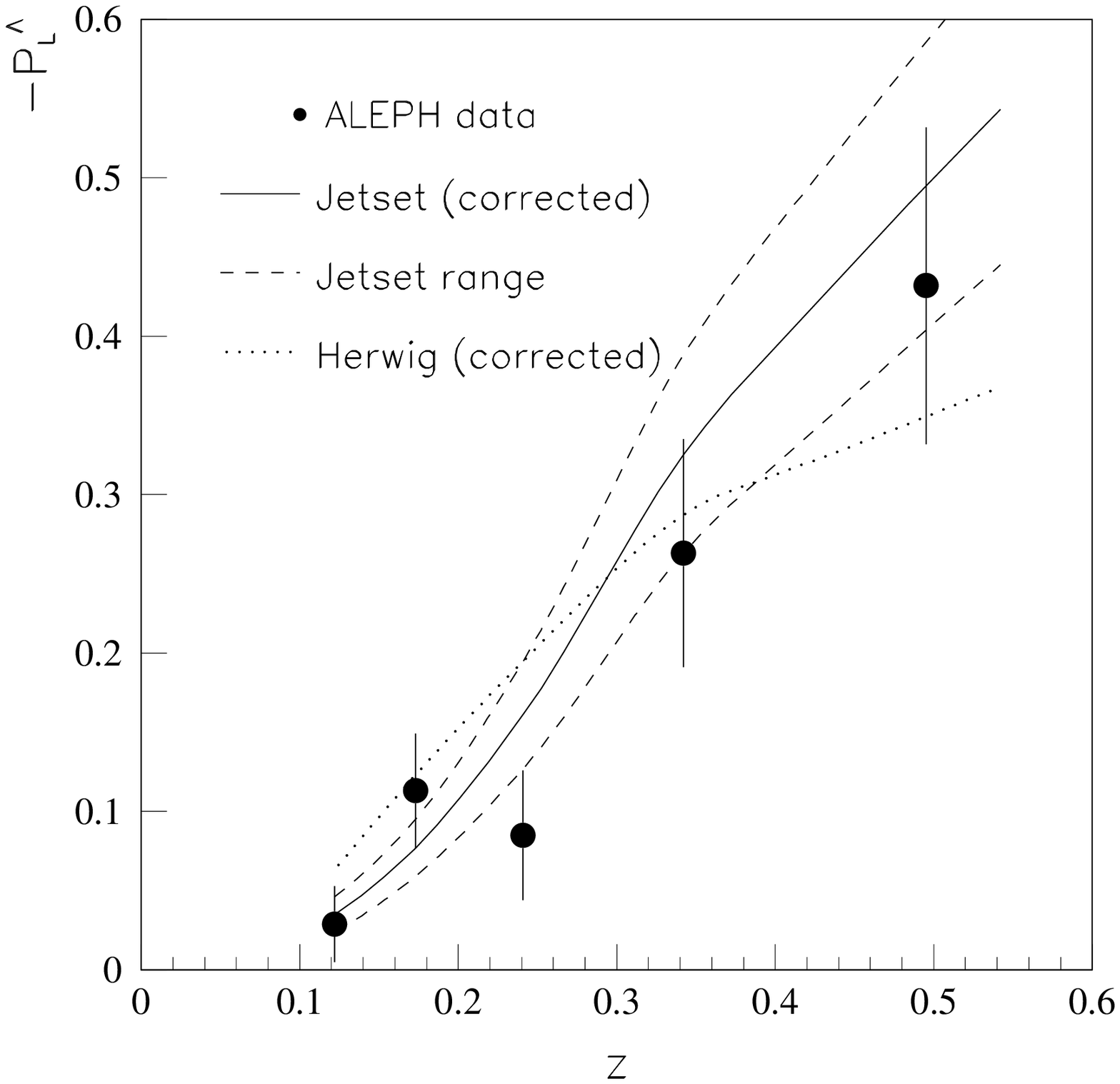,width=8cm}
    {\small Figure 4: ALEPH result for longitudinal polarization
        of $\Lambda $'s in jets.}
\end{wrapfigure}

Any non-zero results in this area are of importance, since they
can be used as an analyzer of parton polarization, for example in
DIS.

\section{Where next?}

I see at least three areas where progress can reasonably be
expected.  When planning experiments, it is important to attempt
to anticipate these areas, for otherwise the design of
experiments to be performed up to a decade ahead will be tied to
the current state of theory rather than to the state of theory
when the experiments are performed.
\begin{itemize}

\item
    In the short term it is important to get the errors in both
    theoretical calculations and in QCD phenomenology under
    better control.  Uncertainty in our knowledge both of
    perturbative quantities (predicted from theory) and of
    non-perturbative quantities (measured from experiment with
    the aid of theoretical formulae) are often the most
    significant source of systematic error in the analysis of
    data.

\item
    In the long term, we need to find ways of treating
    non-perturbative QCD in real time (as opposed to the
    imaginary time that is used in lattice Monte-Carlo
    calculations).  Even without calculations purely from first
    principles, it would be useful to have better discussing
    these phenomena.

\item
    A characteristic phenomenon of non-perturbative QCD is, of
    course, the spontaneous breaking of chiral symmetry.  The
    underlying degree of freedom here is spin, and so we should
    expect polarized scattering to provide important tests of any
    future understanding of non-perturbative QCD.

\end{itemize}

At this conference many of the talks have concerned the
measurement of polarized parton densities.  These are of direct
relevance to the non-perturbative structure of hadrons.  For
example the spin distributions of anti-quarks, of gluons, and of
strange quarks are directly related to the unusual properties of
chiral symmetry and of the proton wave function.  This
particularly applies to the sum rules related to the integrals of
the densities over all $x$.

Moreover a comparison of the transversity distribution of a quark
($\delta q(x)$ or $h_{1}$) and the helicity distribution $\Delta q(x)$ directly
probes relativistic effects in the wave function.  (Normal
non-relativistic quark models have this distributions equal.)
The azimuthal distribution of quarks in the fragmentation of
transversely polarized quarks (the ``sheared jet effect'') [4]
can only exist if chiral symmetry is broken.

Measurements of all these quantities is very likely to be
of great interest in testing any future understanding of
non-perturbative QCD.

\section{Conclusions}

As is well-known, QCD provides many perturbatively based
predictions, but only with the aid of measurable non-perturbative
functions, the parton densities and fragmentation functions, etc.
However, the present accuracy of the predictions leaves much to
be desired; this situation is improving under the stimulus of
experimental data.

The most important question in QCD is to find how to treat it
non-perturbatively in Minkowski space (i.e., with real time).
Since chiral symmetry breaking is an important part of this area,
polarized probes should be important.

\section{Acknowledgements}

This work was supported in part by the U.S. Department of Energy
under grant DE-FG02-90ER-40577. I would like to thank colleagues
for discussions.

\vspace{0.2cm}
\vfill
{\small\begin{description}

\item{[1]}
    See, for example, the reviews in A.H. Mueller (ed.)\
    ``Perturbative QCD" (World Scientific, Singapore, 1989).

\item{[2]}
    J.C. Collins, Nucl.\ Phys.\ {\bf B394} (1993) 169.

\item{[3]}
    R.L. Jaffe, Comments Nucl.\ Part.\ Phys.\ {\bf 14} (1990) 239.

\item{[4]}
    S.J. Brodsky and P. Lepage in Ref.\ [1].

\item{[5]}
    A selection of relevant references is:
    P.V. Landshoff, Phys.\ Rev.\ {\bf D10} (1974) 1024;
    A. Duncan and A.H. Mueller, Phys.\ Rev.\ {\bf D21} (1980)
        1636;
    M.G. Sotiropoulos, Phys.\ Rev.\ {\bf D54} (1996) 808;
    A. Donnachie and P.V. Landshoff, ``The interest of large-$t$
        elastic scattering'', preprint hep-ph/9512397.

\item{[6]}
    E.g., S. Rolli,
    ``Fragmentation Functions Approach in PQCD Fragmentation
    Phenomena'',
    talk given at 31st Rencontres de Moriond: QCD and High-Energy
    Hadronic Interactions, Les Arcs, France, 23-30 Mar 1996.
    E-Print Archive: hep-ph/9607480;
    and references therein.

\item{[7]}
    O. Nachtmann, Nucl.\ Phys.\ {\bf B127} (1977) 314;
    A.V. Efremov, Sov.\ J. Nucl.\ Phys.\ {\bf 28} (1978) 83;
    A.V. Efremov, L. Mankiewicz and N.A. T\"ornqvist, Phys.\
    Lett.\ {\bf B284} (1992) 394.

\item{[8]}
   X. Artru and J.C. Collins,
   Z. Phys.\ C. 69 (1996) 277, and references therein.

\item{[9]}
     R.L. Jaffe, ``Polarized Lambdas in the Current Fragmentation
     Region'', preprint hep-ph/9605456.

\item{[10]}
    ALEPH Collaboration (D. Buskulic et al.),
    Phys.\ Lett.\ {\bf B374}, (1996) 319.

\item{[11]}
    Efremov, these proceedings.

\end{description}}

\end{document}